\documentclass[11pt]{article}
\usepackage{jcapmod}

\usepackage{booktabs}
\usepackage[english]{babel}
\usepackage{amsmath,amssymb,amsbsy,amstext, amsthm, simplewick}
\usepackage{hyperref}
\usepackage{graphicx}
\usepackage{amsfonts}
\usepackage{amssymb}
\usepackage[small]{caption}
\usepackage{upgreek}
\usepackage{wrapfig}
 \usepackage{exscale,relsize}

\usepackage{colortbl}
\definecolor{lightgreen}{cmyk}{0.2, 0, 0.2, 0.2}
\definecolor{lightgray}{cmyk}{0.1,0.2,0,0.1}
\definecolor{lightgray2}{cmyk}{0.1,0.1,0,0.1}

\makeatletter
\newlength{\apb@width}
\newcommand{\autoparbox}[2][c]{\settowidth{\apb@width}{#2}\parbox[#1]{\apb@width}{#2}}

\makeatother

\numberwithin{equation}{section}

\def\beq{\begin{equation}}
\def\eeq{\end{equation}}

\def\bea{\begin{eqnarray}}
\def\eea{\end{eqnarray}}

\def\d{{\rm d}}

\def\cO{{\cal O}}

\def\beq{\begin{equation}}
\def\eeq{\end{equation}}
\def\bea{\begin{eqnarray}}
\def\eea{\end{eqnarray}}

\def\Mp{M_{\rm pl}}
\def\d{{\rm d}}

\DeclareRobustCommand{\SkipTocEntry}[4]{}

\begin{document}

\begin{titlepage}

\setcounter{page}{1} \baselineskip=15.5pt \thispagestyle{empty}

\bigskip\

\vspace{2cm}
\begin{center}

{\fontsize{18}{30}\selectfont  \bf A Field Range Bound \\ \vspace{0.3cm} for General Single-Field Inflation}
\end{center}

\vspace{0.5cm}
\begin{center}
{\fontsize{14}{30}\selectfont   Daniel Baumann$^{\diamondsuit}$ and Daniel Green$^\clubsuit$}
\end{center}


\begin{center}
\vskip 8pt
\textsl{$^\diamondsuit$ DAMTP, Cambridge University, Cambridge, CB3 0WA, UK}

\vskip 7pt
\textsl{$^\clubsuit$ School of Natural Sciences,
 Institute for Advanced Study,
Princeton, NJ 08540, USA}

\end{center}

\vspace{1.2cm}
\hrule \vspace{0.3cm}
{ \noindent \textbf{Abstract} \\[0.2cm]
\noindent 
We explore the consequences of a detection of primordial tensor fluctuations for general single-field models of inflation.  Using the effective theory of inflation, we propose a generalization of the Lyth bound.
Our bound applies to all single-field models with two-derivative kinetic terms for the scalar fluctuations and  
is always stronger than the corresponding bound for slow-roll models.
This shows that non-trivial dynamics can't evade the Lyth bound.
 We also present a weaker, but completely universal bound 
  that holds whenever the Null Energy Condition~(NEC) is satisfied at horizon crossing.}
 \vspace{0.3cm}
 \hrule

\vspace{0.6cm}
\end{titlepage}


\section{Introduction}

The Lyth bound~\cite{Lyth} for single-field slow-roll inflation~\cite{inflation} relates observable tensor modes to a super-Planckian excursion of the canonically-normalized inflaton field, $\Delta \phi > \Mp$. 
Treated as an effective field theory (EFT) with Planck-scale cutoff, 
the inflationary dynamics then becomes sensitive to an infinite number of Planck-suppressed operators~\cite{Lyth,CMBPol}.
For example, the slow-roll potential $V(\phi)$ may receive the following corrections,
\beq
\mathcal{L}  = - \tfrac{1}{2} (\partial_\mu \phi)^2 - V(\phi) \left(1+\sum_{n=1}^{\infty} c_{n} \frac{\phi^n}{\Mp^n} \right) \ . \label{equ:eft}
\eeq
These corrections can be thought of as arising from integrating out Planck-scale degrees of freedom. Generic couplings of these fields to the inflaton $\phi$ result in Wilson coefficients $c_n$ of order one.  The divergence of the series for $\phi > \Mp$ can be interpreted as the breakdown of the effective theory as these heavy fields become massless.
 For super-Planckian fields, every term in~(\ref{equ:eft}) contributes at the same order and will alter the background equally.  
To make sense of slow-roll models with observable tensor modes therefore requires an approximate symmetry that explains why $c_n \ll 1$. 
Moreover, it is desirable that this symmetry is realized in a UV-complete theory such as string theory~\cite{Axion, Axion2} to ensure that it survives any Planck-scale breaking effects~\cite{Kallosh:1995hi}.

For slow-roll inflation, these considerations are well understood.
However, the equivalent statements for more general inflationary models---such as $P(X)$-theories~\cite{ArmendarizPicon:1999rj}, DBI inflation~\cite{Silverstein:2003hf}, ghost inflation~\cite{ArkaniHamed:2003uz} and galileon models~\cite{Burrage:2010cu}---are much less clear.
In this paper, we therefore revisit the Lyth bound for the most general single-field theories of inflation.
Such theories are described in a unified way by the EFT of single-field inflation~\cite{Cheung} (see also~\cite{Senatore:2010wk,Creminelli:2006xe,Baumann:2011su}).  This approach exploits the fact that the inflationary background, $H(t)$, spontaneously breaks time-translation symmetry. Adiabatic fluctuations are then identified with the Goldstone boson, $\pi$, associated with the symmetry breaking. The low-energy EFT of the Goldstone mode can be constructed as a systematic derivative expansion~\cite{Cheung}. 
We are interested in the role of tensor modes in this EFT.
Does a similar Lyth bound exist? What is the relevant ``\hskip 1pt field range\hskip 1pt" in the regime far from slow-roll?

At first sight, the EFT of inflation seems ill-suited for discussing questions about the inflationary background. Being a theory for the inflationary fluctuations, all information about the background is absorbed into the couplings of various operators.  
One may worry that any information about the field range
may be lost by considering only the EFT of fluctuations.
To explain why this is not the case, let us clarify what is special about super-Planckian fields in slow-roll inflation.   
In this case, two related things can happen when $\Delta \phi > \Mp$:  {\it i})~The effective theory breaks down if heavy particles with mass of order $\Mp$ become massless by coupling to $\phi$.  {\it ii})~An infinite number of Planck-suppressed operators contribute equally to physical quantities like the vacuum energy or the masses of particles.
Both of these features 
should be visible in the EFT of inflation:
{\it i})~Planck-mass particles becoming massless surely has a description in the EFT as it must be capable of describing all light fields.  By introducing a time-dependent mass for additional fields, we can capture the same physics. Integrating out the additional fields leads to non-renormalizable operators in the effective theory for the Goldstone mode $\pi$.  {\it ii})~A large field range is distinguished by an infinite number of operators contributing at order one to the generalized slow-roll parameters.  In the theory of the fluctuations, these contributions to the action translate directly into contributions to the mass of $\pi$.  

Formulating a field range bound in the EFT of inflation has certain advantages. First, the concept of field range can be ambiguous when defined in terms of the background field~$\phi$. In particular, far from slow-roll the (naive) ``\hskip1pt field range\hskip 1pt" won't be invariant under field redefinitions. In contrast, our definition of field range in the EFT of inflation will be independent of field redefinitions. Second, the EFT of the Goldstone boson $\pi$ allows a clean interpretation of the energy scales of the problem.
In particular, it shows that two important energy scales characterize all single-field models:
The Hubble scale $H$ corresponds to the energy scale at which curvature fluctuations become time independent.\footnote{We will not consider single-field models with dissipation~\cite{Berera:1995ie, Hall:2003zp, Green:2009ds, Nacir:2011kk}.  In such models, not only is freeze-out modified, but also the scalar~\cite{Berera:1995ie, Hall:2003zp, Green:2009ds, Nacir:2011kk} and tensor~\cite{Cook:2011hg, Senatore:2011sp} modes can be sourced directly.} This is the energy scale that we have access to via CMB observations. The symmetry breaking scale~$\Lambda_b$ defines the energy scale associated with the time variation of the background above which the description in terms of the Goldstone boson $\pi$ may be insufficient. 

Given $H$ and $\Lambda_b$, we will derive a compact and universal form for the power spectrum of curvature perturbations,
\beq
\Delta_\zeta^2 \equiv k^3 P_\zeta \sim \left(\frac{H}{\Lambda_b} \right)^{2+ 2\Delta}\ ,
\eeq
where $\Delta$ is the scaling dimension of $\pi$, such that $\pi \to \lambda^\Delta \pi$ when $\omega \to \lambda \omega$. 
In all examples of interest, the kinetic terms for the Goldstone boson take the form $\Lambda^4 \dot \pi^2$ and the natural size of Planck-suppressed corrections to the mass of $\pi$ is determined by the scale~$\Lambda$. We define ``large-field range" as the regime where an infinite number of operators give order $H$ contributions to the mass of $\pi$.  By relating $\Lambda$ to $\Lambda_b$, we can relate this field range to the tensor-to-scalar ratio~$r$.  This leads to a bound on the field range that is at least as strong as the Lyth bound for slow-roll inflation and typically stronger.  This result applies to virtually all single-field models in the literature.

\vskip 6pt
The layout of this paper is as follows:  In Section~\ref{sec:eft}, we will review the aspects of the effective field theory of inflation relevant to this work.  We will then derive the universal power spectrum and tensor-to-scalar ratio for any single-field model.  Using the Null Energy Condition (NEC), we will prove a completely general upper bound on the tensor-to-scalar ratio.  In Section~\ref{sec:lyth}, we will define the field range using the natural size of Planck-suppressed corrections to the EFT of inflation.  Using this definition, we will prove a field range bound that holds for all models with two-derivative kinetic terms.  We will then show how this bound and the NEC bound combine to make measurable gravity waves in a small-field model a near impossibility.  We will conclude in Section~\ref{sec:conclusions}.

\newpage
\section{Effective Theory of Single-Field Inflation} \label{sec:eft}

\subsection{Adiabatic Fluctuations as Goldstone Bosons}

The fact about inflation that is most relevant to our existence is that it ended.  To do so, inflation requires a physical clock that knows how long the universe has been inflating and can tell it when to stop.  In specific models, this role is typically played by a scalar field with a time-dependent vacuum expectation value (vev).  More formally, inflation spontaneously breaks time-translation symmetry.  As with any spontaneously broken symmetry, this implies the existence of a Goldstone boson. In inflation, the Goldstone boson $\pi$ can be associated with local fluctuations of the clock.  Dynamical gravity gauges the time translations and the Goldstone boson is eaten by the metric, $\zeta = - H \pi$, where $\zeta$ is the comoving curvature perturbation.

In this work, we will focus on the behavior of fluctuations before horizon crossing, $\omega \gtrsim H$.
In this regime, we can ignore the mixing with gravity and focus on the physics of the Goldstone boson alone~\cite{Cheung}. 
Because the theory spontaneously breaks time translations, any time-dependent vev in the complete theory appears in the EFT as an explicit $t$-dependence.  The Goldstone boson restores time translations as an exact symmetry of the action---i.e.~any time dependence should appear in the combination $t+\pi$, such that $\pi \to \pi + 1$ under $t \to t - 1$.  
Given an arbitrary quasi-de Sitter background with $H^2(t) \gg |\dot H|(t)$, we write the action for the Goldstone boson as a derivative expansion in terms of the field $t+\pi$,
\begin{align}\label{action}
S &= \int \d^4 x \sqrt{-g}\, \Big[ \Mp^2 \big[ 3 H^2(t+ \pi) + \dot H(t+\pi) \big] + \Mp^2 \dot H(t+ \pi) \partial_\mu (t + \pi) \partial^\mu (t+\pi) \, + \nonumber \\
 &  \hspace{2.5cm} +\, \sum_{n=2}^{\infty}\tfrac{1}{n!}M_n^4(t+\pi) \left[ (\partial_\mu (t+\pi))^2 + 1\right]^n
+\,  \cdots \Big] \ .
\end{align}
To cancel tadpoles for $\pi$ some of the coefficients were fixed in terms of $H$ and $\dot H$.
The first line in eq.~(\ref{action}) captures all slow-roll models, the second line parameterizes $P(X)$-theories~\cite{ArmendarizPicon:1999rj,Silverstein:2003hf} and $\cdots$ signify terms arising in higher-derivative theories such as ghost inflation~\cite{ArkaniHamed:2003uz} and galileon models~\cite{Burrage:2010cu}.  A priori, all the coefficients in the action may be arbitrary functions of $t+\pi$.  However, scale invariance of the correlation functions requires an additional approximate symmetry under which $t \to t + d$ (with no transformation of $\pi$).  For simplicity, we will take the limit where this is an exact symmetry, so that no explicit functions of $t$ appear in any couplings.  For the leading (slow-roll) terms in the action, this is accomplished by taking the decoupling limit $\Mp \to \infty$ and $\dot H \to 0$ with $\Mp^2 \dot H = const.$ 
For the remaining higher-derivative terms it implies that all couplings are time-independent, e.g.~$M_n \approx const.$

To compute the power spectrum, we need the quadratic terms in the $\pi$-Lagrangian 
\beq \label{quadratic}
\mathcal{L}_{2} =  \underbrace{(- \Mp^2 \dot H + 2 M_2^4)}_{\equiv\, \Lambda^4} \dot \pi^2 + \Mp^2 \dot H (\partial_i \pi)^2 + \tilde M_2^2 (\partial^2 \pi)^2 +\cdots \ ,
\eeq
where $\cdots$ are higher-derivative terms.  At high energies, $\omega \gg H$, the equations of motion derived from (\ref{quadratic}) have approximate flat space solutions $\pi \propto e^{i (\omega t - {\bf k} \cdot {\bf x})}$ and a dispersion relation of the form $\omega = f(k)$.  At $\omega \simeq H$, it is convenient to match to the conserved curvature perturbation $\zeta = - H \pi$.  Typically, a single operator containing spatial derivatives will be dominant at that moment, leading to an approximate dispersion relation of the form $\omega \simeq k^n /\rho^{n-1}$, for some integer~$n$. In general, $n$ need not be an integer, although such cases do not arise from a simple derivative expansion like (\ref{quadratic}).  

Much of the simplifications in the effective theory of inflation arise from a hierarchy of scales that all inflationary models possess.  For example, in slow-roll models, the ratio of the scale at which the time translations are broken ($\dot \phi$) to the freeze-out scale ($H$) is controlled by the amplitude of curvature fluctuations, 
\beq
\Delta_\zeta^2 \sim \frac{H^4}{ \dot \phi^2} \sim \frac{H^4}{\Mp^2 \dot H} \sim 10^{-10}\ .
\eeq
As a result, the Goldstone boson is a reliable description for a wide range of energies.  As we will see in the next two sections, this hierarchy is generic to all inflationary models and has a precise definition arising from the stress tensor.  

\subsection{Stress Tensor during Inflation}

Given an action with a global symmetry, there always exists a conserved current $j^{\mu}$.  When the symmetry is spontaneously broken, this current is still conserved, but the associated charge is no longer well-defined.  For example, if we consider the theory of an abelian Goldstone boson, the current is given by $j^{\mu} = f_{\pi}^2 \partial^\mu \pi + \cdots$~\cite{Weinberg:1996kr}.  At energies below $f_\pi$, the charge ${\cal Q} \equiv \int \d^3 x\,  j^0$ is not well-defined because the integral diverges. This argument identifies the symmetry breaking scale $\Lambda_b$ with $f_\pi$~\cite{Weinberg:1996kr}.   In the case of spontaneous breaking of time translations, the same physics occurs with $j^{\mu} \to T^{\mu 0}$.

\vskip 4pt
It is straightforward to determine the stress tensor of the EFT of inflation by Noether's theorem, 
\begin{align}
T^{\mu}{}_{\nu} &\ =\ -\frac{\delta {\cal L}}{\delta \partial_\mu \pi}\partial_{\nu}(t+\pi) + \delta^{\mu}_{ \nu} {\cal L}  \nonumber \\
&\ =\ \big(2 \Mp^2 \dot H - p_{\pi} \big) \delta^{0 \mu}\partial_{\nu}(t+\pi) + \delta^{i \mu } \delta_{0 \nu} \partial_i \cO + \delta^{\mu}_{ \nu} {\cal L} \ +\ \cdots \ , \label{stresstensor}
\end{align}
where the operator $\cO$ is linear in $\pi$ and $\cdots$ are terms that are at least quadratic in $\pi$. 
We identify the coefficient of the first term in (\ref{stresstensor}) with the canonical momentum,
\beq
\frac{\delta {\cal L}}{\delta\dot \pi} \equiv -2 \Mp^2 \dot H + p_{\pi}\ .
\eeq
Here, we have separated the canonical momentum into a constant contribution $- 2\Mp^2 \dot H$ and an operator $p_{\pi}$ that starts linear in $\pi$ (e.g.~for the slow-roll action $p_\pi = - 2 \Mp^2 \dot H \, \dot \pi$ ).
The constant acts trivially as an operator, while $p_\pi$ satisfies 
\beq
[ p_{\pi}({\bf x},t), \pi({\bf y},t) ]  = -i \delta({\bf x}-{\bf y}) \ .   \label{equ:com}
\eeq
   The second term in (\ref{stresstensor}) is required by conservation of the stress tensor, $\partial_\mu T^{\mu 0} = 0$, i.e.~the term linear in $\pi$ in $T^{00}$ must be matched by a linear term in $T^{i 0}$.
   Moreover, spatial translations are unbroken, so the momentum ${\cal P}^i \equiv \int \d^3 x\, T^{i0} $ has to be well defined at all energies.  Hence, the only way an operator linear in $\pi$ can appear in $T^{i0}$ 
    is as a total derivative, $T^{i 0} \supset \partial^i \cO$ (e.g.~for slow-roll $\partial^i \cO = -2 \Mp^2 \dot H \partial^i \pi$).
   
Although the charge is not well defined in theories with spontaneous symmetry breaking, the commutator of the charge with local operators is still meaningful.  In particular, during inflation, the transformations under time translations are still generated by the Hamiltonian, $\delta \pi(x) =  i [ {\cal H}, \pi(x) ]$, where ${\cal H} \equiv \int \d^3 x \, T^{00}$, even though ${\cal H}$ itself is not well defined at low energies.  Given that we know the transformation properties of the fields, many properties of $T^{00}$ can be determined independently of the form of the action.  In the next section, we will use this to define the scale $\Lambda_b$ at which the time translations are broken.  This works almost in the same way as for the Goldstone boson of an internal symmetry, except that now the theory can be non-relativistic and we have to be careful to define a true `energy' scale~\cite{Baumann:2011su}.
This approach will allow us derive a universal form of the power spectrum $\Delta_{\zeta}^2$ in terms of $\Lambda_b$.

\subsection{Universal Form of the Scalar Power Spectrum}
\label{sec:Ps}

   
 \vskip 4pt
By definition, $T^{00}$ has units of energy over volume, $T^{00} = [\omega] [k]^3$, and $\pi$ has units of time, $\pi = [\omega]^{-1}$.  
Moreover, we will assume that at energies close the Hubble scale, $\omega \sim H$, the Goldstone boson obeys some approximate scaling relation with scaling dimension $\Delta$, i.e.
\beq
\pi \to \lambda^{\Delta} \pi\ , \quad {\rm for} \quad \omega \to \lambda \omega\ . \label{equ:scalingX}
\eeq 
It is convenient to write   
\beq
\pi = \mu^{-1-\Delta}  \omega^{\Delta} \equiv \mu^{-1-\Delta}  \tilde \pi\ ,
\eeq 
where $\omega$ is the energy of $\pi$ and $\mu$ is so far an undetermined energy scale.
This scaling behavior is sufficient to determine the symmetry breaking scale from the stress tensor.
The breaking of time translations occurs due to the linear term\footnote{Naively, there appears to be an additional linear term in (\ref{stresstensor}) coming from $-2 \Mp^2 \dot H \partial_{\nu} \pi$.  However, this piece is cancelled by the linear term in $\delta^{\mu}_{ \nu} {\cal L}$.  Both terms arise from a total derivative in ${\cal L}$ and do not appear in the equations of motion.  It is clear that this cancelation must occur. Under time translations, the Goldstone boson shifts by $\pi \to \pi + 1$, which can be restated as $[T^{00}({\bf x}), \pi({\bf y})] \supset - i \delta({\bf x}-{\bf y})$ .  Any linear term beyond $p_\pi$ would be inconsistent with this charge assignment, unless it has vanishing commutator with $\pi$.} $\delta T^{00} =  p_\pi$, where the conjugate momentum $p_\pi$ has units $[\omega][k]^3$. In order to give the correct scaling in eq.~(\ref{equ:com}),
we define
\beq
\delta T^{00} =  \Lambda_b^{1+ \Delta}\, \tilde p_{\pi} + \cdots\ , \label{equ:T00}
\eeq
where $\tilde p_{\pi}$ scales with energy and momentum as $\omega^{-\Delta} k^3$ and $\Lambda_b$ has units of energy, $\Lambda_b = [\omega]$.  The coefficient in eq.~(\ref{equ:T00}) is the symmetry breaking scale $\Lambda_b$. It controls the term in $\delta T^{00}$ that leads to the divergence in the charge at low energies (like $f_{\pi}$ does for ordinary Goldstone bosons).  
In general, $\Lambda_b^4$ is {\it not} simply the coefficient of the kinetic term\footnote{It is worth remarking that our definition in eq.~(\ref{equ:T00}) did not assume any special form of the action and applies equally to models without conventional kinetic terms~\cite{Baumann:2011su}.} in eq.~(\ref{quadratic}), $\Lambda^4$. 
Furthermore, we have ensured that $\Lambda_b$ is a true `energy' scale whereas $\Lambda^4$ has units of $[\omega][k]^3$ and is hence an `energy density'. 
In practice, one has to use the dispersion relation to relate $\Lambda$ to $\Lambda_b$.  In slow-roll inflation, we find, $\Lambda_b^4 = 2 \Mp^2 |\dot H| = \dot \phi^2$, which is consistent with the intuition that the time dependence of $\phi(t)$ controls where the symmetry is spontaneously broken. In theories with small sound speed, we get $\Lambda_b^4 = 2 \Mp^2 |\dot H| c_s$, while $\Lambda^4 = 2 \Mp^2 |\dot H| c_s^{-2}$~\cite{Baumann:2011su}.  Finally, the commutation relation,
\beq
[p_\pi({\bf x}) , \pi({\bf y})] = (\Lambda_b/\mu)^{1+\Delta} [\tilde p_\pi({\bf x}), \tilde \pi({\bf y})] = -i\delta({\bf x}-{\bf y}) \ ,\eeq implies\footnote{One may instead use $[ \tilde p_{\pi} , \tilde \pi ] = -i\delta({\bf x}-{\bf y})$ as the definition of $\Lambda_b$.  This definition would suffice to eliminate any order-one factors relating $\Lambda_b$ and $\mu$, but leaves undetermined other order-one factors that relate these dimensionful scales to those that appear in the solutions to the equations of motion.  As with all dimensional analysis arguments, we can't determine the dependence on natural numbers.  In principle, there could be accidental factors of 2 or $\pi$ or $e^{-1000}$. All equalities in this section are at the level of dimensionful parameters.  } that $\mu \sim \Lambda_b$.  The mode function near the Hubble scale $H$ therefore behaves as
\beq
\pi \sim \Lambda_b^{-1-\Delta}  \omega^\Delta \ . \label{equ:scaling}
\eeq  
We assume that the modes are in the Bunch-Davies vacuum at high energies and evolve adiabatically as their physical wavelengths are redshifted. It then follows from the covariant conservation equation alone\footnote{In the absence of dissipation, the stress tensor for the inflaton is covariantly conserved, namely
\beq
\nabla_\mu T^{\mu \nu} = (\partial_0 + 3H )T^{00} + \partial_i T^{i0} + H g_{ij} T^{ij} = 0\ .
\eeq 
Linear terms in $\pi$ are conserved using the equations of motion.  When $\omega \gg H$, the equations of motion allow WKB-like solutions with $\omega(t) = f(k(t))$.  The existence of the WKB solutions follows from the observation that the conservation equation is the same as in flat space when we can drop $3 H T^{00}$ and $H g_{ij} T^{ij}$, i.e.~$\omega T^{00} + k_i T^{i0} \sim 0$, for $\omega \gg H$. Under reasonable assumptions, the WKB solutions are valid until $\omega \sim H$, where the contribution to the stress-energy conservation from $3H T^{00}$ is no longer negligible.  For $\omega < H$, we know that $\zeta = - H \pi$ has a constant solution~\cite{Weinberg:2003sw} and therefore freeze-out will occur at~$\omega \sim H$.} 
that the modes freeze out at $\omega \sim H$.
We therefore find
\beq
\zeta = - H \pi \sim  \left. H \frac{\omega^\Delta}{\Lambda_b^{1+\Delta}} \right|_{\omega=H} \sim \Big(\frac{H}{\Lambda_b}\Big)^{1+ \Delta}\ .
\eeq
Hence, we obtain an estimate for the power spectrum of curvature perturbations for theories with the scaling behavior (\ref{equ:scalingX}),
\beq
\Delta_\zeta^2 \equiv k^3 P_\zeta \sim \Big(\frac{H}{\Lambda_b}\Big)^{2+ 2\Delta} \ . \label{equ:DeltaZeta}
\eeq
We combine this with the {model-independent} power spectrum for tensor modes,
\beq
\Delta_h^2 \equiv k^3 P_h \sim \Big(\frac{H}{\Mp} \Big)^2\ ,
\eeq
to get the tensor-to-scalar ratio,
\beq
r  \sim \frac{\Lambda_b^2}{\Mp^2} \left(\frac{\Lambda_b}{H}\right)^{2\Delta}\ . \label{equ:r}
\eeq
This result forms the basis for obtaining a Lyth-like bound on general single-field inflation.

\subsection{Null Energy Condition and Field Range}
\label{sec:NEC}

Given the form of the tensor-to-scalar ratio in (\ref{equ:r}), it might seem that we have the freedom to make $r$ arbitrarily large while keeping the slow-roll parameter $\varepsilon \equiv - \frac{\dot H}{H^2}$ fixed.  In particular, making $\Lambda_b^4 \gg \Mp^2 |\dot H|$ might seem like a promising first step in generating measurable gravity waves in a small-field model.  However, in this section, we will show that in any theory satisfying the Null Energy Condition (NEC), there is an upper bound on the scale $\Lambda_b$ (and hence $r$) for a given value of $\dot H$.  In the next section, we will argue that the bound from the NEC implies that the physically relevant field range can never be made parametrically small compared to the Lyth bound.

\vskip 4pt
Recall that the NEC is the statement that
\beq
T_{\mu \nu} n^\mu n^\nu  \geq 0\ , \label{equ:NEC}
\eeq  
for any null vector $n^\mu n^\nu  g_{\mu \nu} = 0$.
We will follow the logic of Arkani-Hamed et al.~\cite{Adams:2006sv} and apply the NEC to the effective theory of inflation.  
The observed near-Gaussianity of the primordial fluctuations requires that the theory is weakly coupled at $\omega \sim H$, so that we can focus on terms that are linear in operators. 
We found the stress tensor to linear order in fluctuations in eq.~(\ref{stresstensor}).
The NEC becomes 
\beq
\left[- 2\Mp^2 \dot H(1+\dot \pi) + \Lambda_b^{1+ \Delta} \tilde p_{\pi}\right](n^0)^2 + n^{0} n^{i}\partial_i {\cal O} \, \geq\, 0\ .
\eeq
We will drop the term $ -2 \Mp^2 \dot H\, \dot \pi$ because it is suppressed relative to $\Mp^2 \dot H$ by $\Delta_\zeta \sim 10^{-5}$.
Moreover, we are free to choose $n^i$ such that $n^i k_i =0$. We are then left with the following form of the NEC
\beq
- 2\Mp^2 \dot H + \Lambda_b^{1+ \Delta} \tilde p_{\pi} \geq 0\ . \label{equ:NEC2}
\eeq
Because the fluctuations in the second term can take either sign, the NEC is violated if the coefficient $\Lambda_b$ is too large.
Using $\tilde p_\pi \sim \omega^{-\Delta} k^3$ (see \S\ref{sec:Ps}), we evaluate eq.~(\ref{equ:NEC2}) at freeze-out, $\omega \sim H$,
\beq\label{eqn:necbound}
2 \Mp^2 |\dot H| \, \gtrsim \, \Lambda_b^{1+\Delta} H^{3-\Delta} c_p^{-3} \ ,
\eeq
where $c_p \equiv \omega/ k |_{\omega = H}$ is the phase velocity at freeze-out.  
Notice that this is a very conservative bound.
We only used the scaling behavior of $\tilde p_\pi$ at low energies, $\omega \sim H$.  If we extended the scaling of $\tilde p_\pi$ to higher energies, we would get a stronger constraint.  
Combining eq.~(\ref{eqn:necbound}) with eqs.~(\ref{equ:DeltaZeta}) and (\ref{equ:r}), we arrive at a Lyth-like relation
  \beq
 \frac{(2 \Mp^2 |\dot H|)^{1/2} \, \Delta t}{\Mp} \geq \Delta^{1/2}_{\zeta}\cdot \sqrt{r}\, c_p^{-3/2} \,  \Delta N\ . 
  \eeq
In the next section, we will explain how the quantity on the l.h.s.~is related to a generalized notion of the physically relevant field range.  In the case of slow-roll inflation, $2 \Mp^2 |\dot H| = \dot \phi^2$ and the NEC bound is a bound on the conventional field range for the canonically-normalized inflaton.  In that case, the NEC bound is weaker than the Lyth bound by a numerical factor, $\Delta_{\zeta}^{1/2} \sim 10^{-2}$.  The NEC bound will nevertheless be useful as it assumes nothing more than we required to determine the universal form of the scalar power spectrum.  Therefore, it places an absolute upper bound on the tensor-to-scalar ratio.

\section{A Generalized Lyth Bound}
\label{sec:lyth}

\subsection{Field Range in the EFT of Inflation}
\label{sec:range}

The first challenge in defining the concept of a ``\hskip 1pt field range\hskip 1pt" for general models of single-field inflation is to determine a quantity that is invariant under field redefinitions.  In the absence of canonical kinetic terms for the background, the field distance does not have a natural normalization.  We solve this problem by working in the EFT for the fluctuations, as the Goldstone bosons have a natural and unambiguous normalization.  The second challenge is that our definition should be physically meaningful.  In particular, we will demand that our notion of field range controls the natural size of Planck-suppressed corrections to the low-energy action.
Our main diagnostic will be corrections to the mass of $\pi$ (which is massless in the decoupling limit $\dot H \to 0$).
Since these corrections are intimately tied to the size of the slow-roll parameters for the background, they will give us a handle on an appropriate definition of ``\hskip 1pt field range\hskip 1pt".  Because the effective mass of $\pi$ is a physical quantity, our definition will have physical content.  Specifically, when this field range is super-Planckian, an infinite number of independent Planck-suppressed operators contribute masses of order $H$ to the canonically-normalized Goldstone boson $\pi_c$.

\vskip 4pt
{\it Eta problem.} \hskip 6pt As a warm-up, we consider the eta problem. In a slow-roll model, Planck-suppressed corrections of the form $V(\phi) \frac{\phi^2}{\Mp^2}$ contribute to $\eta \equiv \Mp^2 \frac{V''}{V}$ at order one and threaten to end inflation prematurely.  To resolve the eta problem and produce a viable model of inflation, one must explain the absence of these terms.  On the other hand, the EFT of adiabatic fluctuations is valid even when these corrections are included \cite{Creminelli:2006xe}.  However, in this non-inflating FRW universe the fluctuations $\pi$ are massive.
To see this, consider the lowest-order action for the Goldstone boson
\beq
\mathcal{L} = \Mp^2 \dot H (\partial_\mu \pi)^2 - \Mp^2 (3 H^2 + \dot H)\ .
\eeq
In the decoupling limit ($\Mp \to \infty$, $\dot H \to 0$, with $\Mp^2 \dot H = const.$) the Goldstone is exactly massless.
In this limit, the mass for $\pi$ is protected by a global shift symmetry, $\pi \to \pi + d$, with no associated time translation.  The eta problem refers to the fact that the symmetry is broken by Planck-suppressed operators that don't vanish in the decoupling limit,
\beq
3 \Mp^2 H^2 \, \to\, 3\Mp^2 H^2 \left(1 + c \, \frac{\Mp^2 \dot H}{\Mp^2} (t+ \pi)^2 \right)\ . \label{equ:HSR}
\eeq
This generates a mass for the canonically-normalized field $\pi_c^2 \equiv 2\Mp^2 |\dot H| \pi^2$ of the form $c\hskip 1pt H^2 \pi_c^2$.  In slow-roll inflation, $\Mp^2 \dot H = \frac{1}{2} \dot \phi^2$ and this statement is identical to the usual eta problem.

This logic generalizes straightforwardly to a broader class of single-field models. 
Consider theories with two-derivative kinetic terms 
\beq
\label{equ:disp}
{\cal L} = \Lambda^4 \dot \pi^2 + \cdots \equiv \tfrac{1}{2} \dot \pi_c^2 + \cdots \ .
\eeq
This captures slow-roll inflation~\cite{inflation}, $\Lambda^4 =  \Mp^2 |\dot H|$, $P(X)$-theories~\cite{ArmendarizPicon:1999rj,Silverstein:2003hf}, $\Lambda^4 =  \Mp^2 |\dot H| c_s^{-2}$, ghost inflation~\cite{ArkaniHamed:2003uz}, $\Lambda^4 = M^4$, and galileon inflation~\cite{Burrage:2010cu}, $\Lambda^4 =  \Mp^2 |\dot H| c_s^{-2}$. 
To avoid superluminal propagation of $\pi$, we require $\Lambda^4 \geq \Mp^2 |\dot H|$ \cite{Cheung}.
Like in eq.~(\ref{equ:HSR}), we can have Planck-suppressed corrections to the energy density,
\beq\label{equ:generalcorrections}
3\Mp^2 H^2 \, \to\, 3 \Mp^2 H^2 \left(1 + c\, \frac{\Lambda^4}{\Mp^2} (t+ \pi)^2 \right)\ .
\eeq
Without knowing anything about the Wilson coefficient $c$, the choice of scale $\Lambda$ may seem arbitrary.  It will be important for the rest of the paper that $c \sim \cO(1)$ is generic.

The appearance of the scale $\Lambda$ in (\ref{equ:generalcorrections}) is a reflection of the fact that $\pi_c$ couples to gravity like any other light field.  Consider, for example, the coupling of $\pi_c$ to the linearize metric perturbation $h_{ij} = \Mp \delta g_{\ij}$.  From the kinetic term for $\pi_c$, we find the coupling
\beq
{\cal L}_{\rm int} = \frac{1}{8 \Mp^2} h_{ij} h^{ij} \dot \pi_c^2  =\frac{\Lambda^4}{4 \Mp^2} h_{ij} h^{ij} \dot \pi^2 \ .
\eeq
As with all gravitational interactions, this becomes strongly coupled when $\omega \sim \Mp$.  Whatever physics UV completes gravity at theses scales should couple to $\pi_c^2$ with order-one couplings in order to regulate the growth of scattering amplitudes at these high energies. 
Since Planck-scale physics is not expected to respect any global symmetries~\cite{Kallosh:1995hi}, there is no reason it would not couple to $\pi_c$ directly.  Integrating out this new physics generates the term in (\ref{equ:generalcorrections}) with $c \sim {\cal O}(1)$.
The mass term for the canonically-normalized Goldstone boson then is
  \beq
\Mp^2  H^2 \frac{\Lambda^4}{\Mp^2} \pi^2 \sim H^2 \pi_c^2\ .
  \eeq

\vskip 4pt
{\it Large field ranges.} \hskip 6pt In slow-roll models of large-field inflation, we should worry about corrections from an infinite number of Planck-suppressed operators, see eq.~(\ref{equ:eft}).
In the EFT of inflation, these terms take the form
\beq
3 \Mp H^2 \, \to\, 3 \Mp^2 H^2 \left(1 + \sum^{\infty}_{n=1}c_n \Big(\frac{\Mp^2 \dot H}{\Mp^2} (t+ \pi)^{2} \Big)^n \right) \ .
\eeq
The issue is the same as before: when $|\dot H| t^2 > 1$, we have to check that every single Wilson coefficient is small, $c_n \ll 1$.  The equivalent situation arises in theories like eq.~(\ref{equ:disp}) if we make the replacement
\beq
3 \Mp^2 H^2 \to 3 \Mp^2 H^2 \left(1  + \sum^{\infty}_{n=1}c_n \Big(\frac{\Lambda^4}{\Mp^2} (t+ \pi)^{2} \Big)^n \right) \ . \label{equ:Lcor}
\eeq
The contribution to the mass of $\pi_c$ from any term in eq.~(\ref{equ:Lcor}) is given by 
\beq
c_n\,  \Mp^2 H^2 \Big(\frac{\Lambda^4}{\Mp^2}\Big)^n t^{2n-2} \pi^2 =c_n\, H^2  \Big(\frac{\Lambda^4 t^2}{\Mp^2} \Big)^{n-1} \pi_c^2 \ .
\eeq
Hence, if $\Lambda^4 t^2 > \Mp^2$ there are an infinite number of terms that contribute dangerously large masses to $\pi_c$.
This motivates us to define ``\hskip 1pt  large field range\hskip 1pt" as
\beq
\frac{\Lambda^2 \Delta t }{ \Mp}  > 1\ , \label{equ:large}
\eeq
where $\Delta t = t_f -t_i$ parameterizes the time interval between horizon exit of CMB scales and the end of inflation.

\subsection{Lyth Bound for Single-Field Inflation}

With a definition of field range in hand, we now wish to relate it to the size of the tensor-to-scalar ratio.  
We have written both quantities in terms of physical scales of the EFT of inflation.  All that remains is to find the relation between these scales, so that we can recast $r$ in terms of the field range. 

\vskip 4pt
To relate $\Lambda$ to $\Lambda_b$ and hence $r$, we require knowledge of the dispersion relation.
Recall that we are considering theories with an approximate scaling symmetry: $\omega \to \lambda \omega$ and $\pi \to \lambda^\Delta \pi$, valid near $\omega \sim H$.
This implies the following dispersion, 
\beq
\omega = k^n/\rho^{n-1}\ , \qquad {\rm where} \quad n = \frac{3}{1 + 2 \Delta} \ . \label{equ:w}
\eeq
Here, $\Delta$ is determined by $n$ since we have assumed that the dominant kinetic term is $\dot \pi_c^2$.  For the special case $n=1$, we define $\omega = c_s k$. 
The leading contribution to the time-time component of the stress tensor is 
\beq
\delta T^{00} = 2\Lambda^4 \dot \pi  \equiv p_\pi = \Lambda_b^{1+\Delta} \tilde p_\pi\ ,
\eeq  
where $\Lambda^4 = [\omega] [k]^3$. As their appearance in the stress tensor suggests, $\Lambda$ and $\Lambda_b$ refer to the same physical scale, just written in different units.  To determine $\Lambda_b^4=[\omega]^4$, we use the dispersion relation (\ref{equ:w}),
\beq\label{equ:LtoL}
\Lambda_b = \Lambda \left( \frac{\Lambda}{\rho} \right)^{\frac{3(n-1)}{n+3}}\ .
\eeq
Using (\ref{equ:LtoL}), the tensor-to scalar ratio (\ref{equ:r}) can be written in terms of $\Lambda$,
\beq
r= \Big( \frac{H}{\rho} \Big)^{3-3/n} \frac{\Lambda^4}{\Mp^2 H^2} \equiv c_p^{3} \hskip 1pt \frac{\Lambda^4}{\Mp^2 H^2} \ ,
\eeq
where, as before, $c_p \equiv \omega/ k |_{\omega = H}$ is the phase velocity at horizon crossing.
We get the following relation
\beq
\frac{\Lambda^2 \Delta t }{ \Mp} \sim \sqrt{r}\,  c_p^{-3/2} \, \Delta N\ . \label{equ:NewLyth}
\eeq
This should be compared with the original Lyth bound for slow-roll inflation,
\beq
\frac{\Delta \phi}{\Mp} \sim \sqrt{r}\, \Delta N\ . \label{equ:OldLyth}
\eeq
Of course, the two agree in the slow-roll limit where $c_p = 1$.  
Because the Goldstone boson is massless (i.e.~$\omega \to 0$ as $ k \to 0$), it follows that if $c_p(\omega_0) >1$ for some energy $\omega_0$ then the group velocity $c_g(\omega_1) \equiv d\omega/ dk |_{\omega_1} > 1$ for some other energy $0< \omega_1 < \omega_0$.
For massless particles, having $c_p > 1$ anywhere therefore implies
superluminal propagation, $c_g > 1$, somewhere.
To avoid potential pathologies, we require $c_p(H) < 1$.
This implies that the bound (\ref{equ:NewLyth}) is always stronger than the original Lyth bound (\ref{equ:OldLyth}).

\subsection{Implications for Explicit Models}

We have argued that $\Lambda^2 \Delta t$ is the natural definition of the physically relevant field range for theories where the fluctuations are governed by the kinetic term $\Lambda^4 \dot \pi^2$.  We were then able to derive a bound on this quantity that is at least as strong as the Lyth bound.  Two obvious questions that one might like to address are: (1) how do we understand this result in explicit models, and (2) can we relax the condition on the form of the kinetic term? 

\vskip 4pt
{\it Corrections to $P(X)$-theories.} \hskip 6pt 
Many single-field models described by the EFT of inflation arise from expanding around so-called $P(X)$-theories~\cite{ArmendarizPicon:1999rj,Silverstein:2003hf}, with Lagrangian
\beq
{\cal L} = P(X,\phi) - V(\phi) \ ,
\eeq
where $X \equiv - (\partial_\mu \phi )^2$ and $P(X, \phi)$ is some function to be specified.  
The naive Lyth bound for the inflaton field $\phi$ is~\cite{Baumann:2006cd}
\beq
\frac{\Delta \phi}{\Mp} \sim \sqrt{\frac{r}{c_s P_{,X}}}\, \Delta N \ , \label{equ:naive}
\eeq
where $P_{,X} \equiv \partial_X P$ and
\beq
c_s^{2} \equiv \frac{P_{,X}}{P_{,X} + 2 X P_{,XX}} \, \le\,  1\ .
\eeq
Eq.~(\ref{equ:naive}) suggests that large tensors could arise without producing super-Planckian vev's if we could make $P_{,X} \gg 1$ for fixed $c_s$.
In fact, various previous works have considered this possibility. 
Here, we argue that these attempts to get around the Lyth bound are somewhat misguided. 

First, we should note that $P_{,X} \gg 1$ implies that the kinetic term of the theory is far from canonical. It is therefore not clear anymore that $\Delta \phi$ is the relevant field range. 
This is precisely the regime where the effective theory of the fluctuations is most useful.
Given an inflationary solution, the theory for the fluctuations is described, as usual, by expanding in $\phi({\bf x},t) = \bar \phi(t) + \dot{\bar\phi}(t) \pi({\bf x},t)$.  Taylor expanding $P(X, \phi)$ around such a background, one finds that 
\beq
\Lambda^4 = \bar X P_{,\bar X} + 2 \bar X^2 P_{,\bar X\bar X} = \frac{\bar X P_{,\bar X}}{c_s^2}\ .
\eeq
For $P_{,\bar X} \gg 1$, this implies 
\beq
\Lambda^2 \Delta t = \frac{\sqrt{P_{, \bar X}}}{c_s} \, \dot{\bar \phi}\, \Delta t \ \gg\ \Delta \phi\ .
\eeq
Large $P_{, \bar X}$ hence leads to large $\Lambda^2 \Delta t$ even if $\Delta \phi$ is small. We don't win by making $P_{,\bar X}$ large. Corrections like in eq.~(\ref{equ:Lcor}) are still a concern.

 What do these corrections correspond to in the theory of the background?
 In slow-roll inflation, the corrections we considered were 
 \beq
\Delta {\cal L} \ =\ - V(\phi) \frac{\phi^2}{\Mp^2}\ . \label{equ:DDL}
 \eeq  
 However, one should include all possible corrections, and if $P_{,X} \gg 1$, the corrections we were proposing in \S\ref{sec:range} are much larger than the correction in (\ref{equ:DDL}). 
 To identify these corrections in $P(X)$-theories, consider deforming the action as follows
\beq\label{equ:pxcorrection}
\Delta {\cal L} \ =\ P\big(X - V(\phi) \tfrac{\phi^2}{\Mp^2} , \phi \big) - V(\phi) \ =\ P(X, \phi) - V(\phi) \Big( 1 + P_{,X} \frac{\phi^2}{\Mp^2} \Big) + \cdots \ .
\eeq
We see that introducing corrections directly to $P(X, \phi)$ reproduces the corrections proposed in the effective theory.\footnote{We can also see these corrections appearing within the context of the effective theory by introducing supersymmetry (SUSY).  As SUSY only protects the mass down to the Hubble scale, we typically find that supergravity corrections give a mass of order $H$, unless we include a shift symmetry for the inflaton.  For theories with $c_s \ll 1$, the supergravity corrections match those proposed here with $\Lambda^4 = \Mp^2 |\dot H| c_s^{-2}$~\cite{Baumann:2011nk}. }   
For $P_{,X} \gg 1$, these corrections are enhanced relative to the correction to the potential.
For slow-roll models, $P_{,X} = 1$, the correction in (\ref{equ:pxcorrection}) is identical to the correction to the potential (\ref{equ:DDL}).

\vskip 4pt
{\it Arbitrary kinetic terms and the NEC.} \hskip 6pt The results derived in this section were, so far, restricted to models with two-derivative kinetic terms.  If we relaxed this condition, is it possible to generate gravity waves without having to worry about large corrections? 
Recall that, in \S\ref{sec:NEC}, we derived the following bound for theories satisfying the
 NEC  at horizon crossing,  
\beq
\frac{(2 \Mp^2 |\dot H|)^{1/2} \, \Delta t}{\Mp} \geq \Delta^{1/2}_{\zeta}\cdot \sqrt{r}\, c_p^{-3/2} \,  \Delta N\ . 
\eeq
This result made no assumptions about the form of the action.  We can now translate this result into a statement about the minimal corrections to the action.  First, we note that even if the theory at horizon crossing is not controlled by the two-derivative kinetic term $\Lambda^4 \dot \pi^2$, the absence of superluminal modes still requires that it is present with a coefficient $\Lambda^4 \geq \Mp^2 |\dot H|$.  The corrections to the action at high energies should not depend on the specific operator that dominates at horizon crossing, so we expect that corrections should be at least as large as those expected from the canonical kinetic term.  Therefore, whatever definition of field range ``\hskip 1pt$\Delta \phi$\hskip 1pt" is appropriate for these more general models, it should satisfy
\beq\label{equ:necbound2}
\frac{ ``\hskip 1pt\Delta \phi\hskip 1pt " }{\Mp} \geq \Delta^{1/2}_{\zeta}\cdot \sqrt{r}\, c_p^{-3/2} \,  \Delta N\ .
\eeq
Although this bound is weaker than (\ref{equ:NewLyth}) by the numerical factor $\Delta_\zeta^{1/2} \sim 10^{-2}$, it presents little room for engineering controlled models in field theory with measurable gravity waves.

\vskip 4pt
{\it Desensitizing inflation.} \hskip 6pt  
Corrections to the inflaton action arise from integrating out massive degrees of freedom at the Planck scale.  On the other hand, the scale at which modes freeze-out is the inflationary Hubble scale, $H \sim \Delta_{\zeta} \sqrt{r} \Mp$.  Therefore, even for a measurable tensor amplitude, $r \gtrsim 0.01$, the physics of inflation happens at an energy scale that is five orders of magnitude below the scale where the quantum gravity corrections are being generated.  
Because of this large ratio of scales, it is conceivable that these corrections are absent at low energies as the result of significant RG flow.  In the context of the eta problem, it was shown that changing the dimensions of operators near the Planck scale can reduced these corrections to an acceptable size \cite{Baumann:2010ys}.  One might wonder if a similar mechanism could explain the absence of large corrections for models producing measurable gravity waves.

As a concrete example, consider the following two-field action~\cite{Baumann:2011su}
\beq
\mathcal{L} =  -\tfrac{1}{2} (\partial_\mu \pi_c)^2 - \tfrac{1}{2} \left[(\partial_\mu \sigma)^2 + \mu^2 \sigma^2\right] + \rho \dot \pi_c \sigma + \cdots \ ,
\eeq
where $\pi_c^2 \equiv 2 \Mp^2 |\dot H| \pi^2$ and $\cdots$ stands for are all operators that are not quadratic in the fluctuations.  At high energies, $\omega \gg \rho$, the theory is well described by two decoupled scalar fields, $\pi_c$ and $\sigma$.  At energies $\omega < \rho$, it becomes a single-field model governed by a non-relativistic kinetic term.  At very low energies, $\omega < \mu^2/ \rho$, it becomes a model with small speed of sound $c_s \simeq \mu/\rho$.  Therefore, at high energies, the kinetic term is $ \Mp^2 |\dot H|\, \dot \pi^2 \equiv \Lambda_{\mathsmaller{\rm UV}}^4\hskip 1pt \dot \pi^2$, while at low energies, it is $ \Mp^2 |\dot H| c_s^{-2} \, \dot \pi^2 \equiv \Lambda_{\mathsmaller{ \rm IR}}^4 \hskip 1pt\dot \pi^2$.  Due to the significant amount of RG flow, we have increased $\Lambda^4 =  \Lambda_{\mathsmaller{ \rm IR}}^4$ at horizon crossing. From the bottom-up the corrections therefore look worse than they are from the top-down. 

However, this model still fails to achieve measurable gravity waves with a small field range.  Using $\Lambda_{\mathsmaller{\rm UV}}$ in the definition of the field range and noting that the bound in (\ref{equ:NewLyth}) applies to $\Lambda_{\mathsmaller{\rm IR}}$ (with $c_p = c_s$), we find 
\beq
\frac{\Lambda_{\mathsmaller{\rm UV}}^2 \Delta t}{\Mp} \equiv \frac{(\Mp^2 |\dot H|)^{1/2} \Delta t}{\Mp} = \frac{c_s \Lambda_{\mathsmaller{\rm IR}}^2 \Delta t}{\Mp} \geq \sqrt{r}\, c_s^{-1/2}\, \Delta N \ .
\eeq
  Despite achieving $\Lambda_{\mathsmaller{\rm IR}} \gg \Lambda_{\mathsmaller{\rm UV}}$, the field range bound is still stronger than the Lyth bound.

The obstacle to achieving small field ranges is not restricted to this example.  Any proposed mechanism,  even with this type of RG flow, is ultimately limited by the NEC bound (\ref{equ:necbound2}).  
The constraint from the NEC depends only on the values of $\Mp^2 \dot H$ and $\Lambda_b$ at horizon crossing.  If $\Mp^2 \dot H$ is independent of scale, then the bound (\ref{equ:necbound2}) provides a lower limit on Planck-scale corrections, no matter how $\Lambda$ evolves under RG flow.  The only way to weaken the bound is for $\Mp^2 \dot H$ to be larger in the IR than in the UV.  However, to leading order, $\Mp^2 \dot H$ is simply a constant in the EFT of inflation and is not altered by RG flow.  Any attempt to modify this coefficient must take place at the level of the background and is beyond the scope of the EFT and this work (but see~\cite{BenDayan:2009kv, Hotchkiss:2011gz}).

\section{Conclusions}
\label{sec:conclusions}

A stochastic background of tensor modes is arguably one of the most robust predictions of inflation. 
Realistically, a tensor signal will be observable in CMB polarization if the tensor-to-scalar ratio $r \equiv P_t/P_s$ is bigger than $0.01$~\cite{CMBPol}.
Remarkably, this level of gravity waves, seems to be tied to Planck-scale physics.
Under the restrictive assumption of {\it slow-roll} inflation,
Lyth~\cite{Lyth} showed that $r > 0.01$ corresponds to super-Planckian evolution of the inflaton field,
\beq
\frac{\Delta \phi}{\Mp} \sim \sqrt{\frac{r}{0.01}} \ . \label{equ:lyth}
\eeq
Having the field traverse a distance larger than the cutoff provides a challenge for a controlled effective field theory description~\cite{CMBPol}. On the other hand, it provides an opportunity for UV-complete treatments of inflation such as string theory realizations of large-field inflation~\cite{Axion, Axion2}.

However, maybe assuming slow-roll is too limiting.
We would like to be able to interpret future data without making strong theoretical assumptions.
In this paper, we have therefore widened the scope of the Lyth bound.
To achieve this, we employed the
EFT of inflation~\cite{Cheung} which describes all possible single-field theories in a single, unified framework.
We showed that the power spectrum of scalar fluctuations can be expressed in a simple, unified form,
in terms of the
Hubble scale $H$, the symmetry breaking scale $\Lambda_b$ and the scaling dimension $\Delta$ of the fluctuations (see \S\ref{sec:Ps}).
The EFT of inflation also allowed us to give a natural definition of the field range $``\hskip 1pt \Delta \phi \hskip 1pt"$ which determines the relevance of Planck-suppressed corrections (see \S\ref{sec:range}).
These arguments culminated in the
generalized Lyth bound
\beq
\frac{``\hskip 1pt \Delta \phi \hskip 1pt"}{\Mp} \sim c_p^{-3/2} \cdot \sqrt{\frac{r}{0.01}} \ ,
\eeq
where $c_p$ is the phase velocity at horizon crossing, $\omega \sim H$.
Since we require $c_p \le 1$ to avoid superluminal modes, our bound is always stronger than the Lyth bound~(\ref{equ:lyth}).
Our result shows that non-trivial dynamics can't evade the Lyth bound.
The UV-sensitivity of observable gravity waves is a universal phenomenon and not special to slow-roll inflation.

\acknowledgments
We are grateful to Nima Arkani-Hamed, Liam McAllister and Rafael Porto for discussions.
D.B.~gratefully acknowledges support from a Starting Grant of the European Research Council (ERC STG grant 279617) and partial support from STFC under grant ST/FOO2998/1.
The research of D.G.~is supported by the DOE under grant number DE-FG02-90ER40542 and the Martin A.~and Helen Chooljian Membership at the Institute for Advanced Study. 

\newpage
 \begingroup\raggedright\endgroup

\end{document}